\documentclass[%
 reprint,
superscriptaddress,
 amsmath,amssymb,
 aps,
]{revtex4-2}

\usepackage{graphicx}% Include figure files
\usepackage{dcolumn}% Align table columns on decimal point
\usepackage{bm}% bold math
\usepackage{color}
\usepackage{hyperref}% add hypertext capabilities
\begin{document}

\preprint{}

\title{The Origin of Two-dimensional Electron Gas in Zn$_{1-x}$Mg$_x$O/ZnO Heterostructures}

\author{Xiang-Hong Chen}
\affiliation{Tianjin Key Laboratory of Low Dimensional Materials Physics and Preparing Technology, Department of Physics, Tianjin University, Tianjin 300354, China}
\author{Dong-Yu Hou}
\affiliation{Tianjin Key Laboratory of Low Dimensional Materials Physics and Preparing Technology, Department of Physics, Tianjin University, Tianjin 300354, China}
\author{Zhi-Xin Hu}
\affiliation{Center for Joint Quantum Studies and Department of Physics, Tianjin University, Tianjin 300354, China}
\author{Kuang-Hong Gao}
\email[Corresponding author, e-mail: ]{khgao@tju.edu.cn}
\affiliation{Tianjin Key Laboratory of Low Dimensional Materials Physics and Preparing Technology, Department of Physics, Tianjin University, Tianjin 300354, China}
\author{Zhi-Qing Li}
\email[Corresponding author, e-mail: ]{zhiqingli@tju.edu.cn}
\affiliation{Tianjin Key Laboratory of Low Dimensional Materials Physics and Preparing Technology, Department of Physics, Tianjin University, Tianjin 300354, China}

\date{\today}% It is always \today, today,

\begin{abstract}
Although the two-dimensional electron gas (2DEG) in (001) Zn$_{1-x}$Mg$_x$O/ZnO heterostructures has been discovered for about twenty years, the origin of the 2DEG is still inconclusive. In the present letter, the formation mechanisms of 2DEG near the interfaces of (001) Zn$_{1-x}$Mg$_x$O/ZnO heterostructures were investigated via the first-principles calculations method. It is found that the polarity discontinuity near the interface can neither lead to the formation of 2DEG in devices with thick Zn$_{1-x}$Mg$_{x}$O layers nor in devices with thin Zn$_{1-x}$Mg$_{x}$O layers. For the heterostructure with thick Zn$_{1-x}$Mg$_{x}$O layers, the oxygen vacancies near the interface introduce a defect band in the band gap, and the top of the defect band overlaps with the bottom of the conduction band, leading to the formation of the 2DEG near the interface of the device. For the heterostructure with thin Zn$_{1-x}$Mg$_{x}$O layers, the absorption of hydrogen atoms, oxygen atoms, or OH groups on the surface of Zn$_{1-x}$Mg$_{x}$O film plays a key role for the formation of 2DEG in the device. Our results manifest the sources of 2DEGs in Zn$_{1-x}$Mg$_x$O/ZnO heterostructures on the electronic structure level.
\end{abstract}

\maketitle

%\tableofcontents

Since the discovery of two-dimensional electron gas (2DEG) at the interface of LaAlO$_3$/SrTiO$_3$ heterojunction~\cite{Ohtomo_Nature2004}, 2DEG has been found at various oxide heterostructures, such as Zn$_{1-x}$Mg$_x$O/ZnO~\cite{Koike_JJAP2004,Tsukazaki_Sci2007,Tsukazaki_NatMater2010}, Al$_2$O$_3$/SrTiO$_3$~\cite{Lee_Nanoscale2013,Schutz_PRB2015}, EuO/KTaO$_3$~\cite{Zhang_PRB2018}, (Al$_x$Ga$_{1-x}$)$_2$O$_3$/Ga$_2$O$_3$~\cite{Zhang_APL2018} and LaAlO$_3$/KTaO$_3$~\cite{Chen_Sci2021}. The 2DEG at oxide heterostructures not only provides a platform for fundamental research, but also promotes the development of novel all-oxide electronic devices. Among these oxide heterostructures, Zn$_{1-x}$Mg$_x$O/ZnO heterostructures are particularly attractive due to their ultra-high Hall mobility (up to $10^6$\,cm$^2$V$^{-1}$s$^{-1}$ at low temperature~\cite{Falson_SciRep2016}). However, the origin of the 2DEG at the Zn$_{1-x}$Mg$_x$O/ZnO interface is still unclear. Researchers only empirically attribute it to the polar discontinuity~\cite{Chin_JAP2010,Ye_SciRep2012,Makino_PRB2013,Tampo_APL2008}: since Zn$_{1-x}$Mg$_x$O ($0<x<0.6$) and ZnO have different spontaneous polarization, the polarization at the interface is discontinuous after they form heterojunctions. This discontinuity causes a large number of bound charges to be generated at the heterointerface, creating a built-in electric field throughout the heterostructure. This field drives electrons toward the interface to form 2DEG. In contrast, some researchers believe that the 2DEG at the Zn$_{1-x}$Mg$_x$O/ZnO interface originates from the donor on the Zn$_{1-x}$Mg$_x$O surface~\cite{Tampo_APL2009,Ye_APL2010}. Experimentally, 2DEG can also be formed when the thickness of Zn$_{1-x}$Mg$_x$O layer is greater than 300\,nm in Zn$_{1-x}$Mg$_x$O/ZnO heterostructures~\cite{Falson_APE2011,Hwang_ACS_Mater_Inter2017,Imasaka_APL2014,Akasaka_JJAP2011}. There would be no internal potential gradient in the aforementioned heterostructure with thick Zn$_{1-x}$Mg$_x$O layer, and the contribution of surface donors to 2DEG could also be negligible~\cite{Yoo_NPJ_Comput2021,Brown_NatCommun2014}. Thus the formation of 2DEG in this case cannot be explained by the mechanisms mentioned above. On the whole, the origin of 2DEG at Zn$_{1-x}$Mg$_x$O/ZnO heterointerface needs to be further studied. In this letter, the origin of 2DEG at Zn$_{1-x}$Mg$_x$O/ZnO heterointerface is studied from the perspective of microscopic electronic structures by first-principles calculations. Interestingly, it is found that the polar discontinuity mechanism is not responsible for the formation of the 2DEG. For the heterostructures with thick Zn$_{1-x}$Mg$_x$O layers, 2DEG mainly arises from oxygen vacancies, while the 2DEG originates from surface adsorption for heterostructures with thin Zn$_{1-x}$Mg$_x$O layers.

\begin{figure}[htpb]
\includegraphics[scale=0.9]{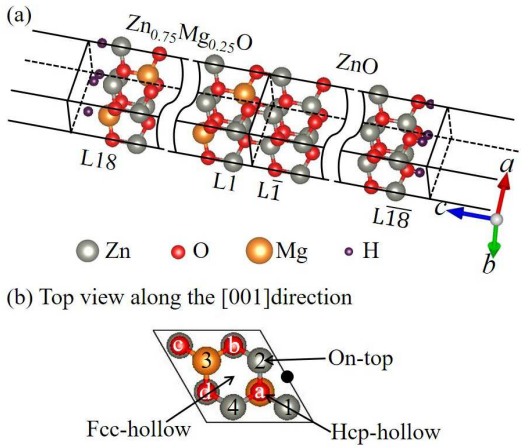}
\caption{(a) Schematic geometrical structure of Zn$_{1-x}$Mg$_{x}$O/ZnO ($x$=0.25 and 0.50) heterostructure with two pseudo-H-passivated surfaces. (b) The top view of the heterostructure along the [001] direction. Here the ``On-top, Fcc-hollow, and Hcp-hollow" are the adsorption sites for exotic atoms or groups.}\label{geome-struc-ZnMgo-ZnO}
\end{figure}

Considering that the 2DEG can be formed in Zn$_{1-x}$Mg$_x$O/ZnO heterostructures with both thick ($\sim$100 to 500\,nm)~\cite{Huang_PRB2005,Tampo_APL2007,Kresse_PRL2003,Dulub_PRL2003,Valtiner_PEL2009} and thin Zn$_{1-x}$Mg$_x$O layers ($\sim$10 to 30\,nm)~\cite{Tampo_APL2008,Tampo_APL2009,Ye_APL2010} experimentally, we construct the configurations as follows. For Zn$_{1-x}$Mg$_x$O/ZnO heterostructures with thick Zn$_{1-x}$Mg$_x$O layers, we passivated the oxygen terminal of ZnO slab and the Zn-Mg terminal of Zn$_{1-x}$Mg$_x$O slab by pseudo-H atoms with fractional charges. ZnO slab with passivated oxygen terminal can be used to simulate ZnO substrate, and the charge of H is taken as $0.48e$ with $e$ being the elementary charge~\cite{Yoo_NPJ_Comput2021}. The charge of the pseudo-H atoms in the passivated Zn-Mg terminal is taken as $1.52e$~\cite{Yoo_NPJ_Comput2021}. After passivation, the pseudo-H atoms not only saturate the surface dangling bonds but also make the passivated surface and the adjacent atomic layers exhibit bulk properties~\cite{Yoo_NPJ_Comput2021,Huang_PRB2005}. In this case, the Zn$_{1-x}$Mg$_x$O and ZnO slabs can be treated as semi-infinite thick films. Considering the Mg content $x$ can be as high as 0.60 in Zn$_{1-x}$Mg$_x$O/ZnO heterostructures experimentally~\cite{Tampo_APL2007}, we set the Mg content $x$ as 0.25 and 0.50, respectively. For each doping level, the Mg ions are uniformly doped into the ZnO film, which together with the ZnO substrate forms a heterostructure with a clear interface. For the Zn$_{1-x}$Mg$_x$O/ZnO heterostructures with thin Zn$_{1-x}$Mg$_x$O layers, the difference in the configuration is that there is no pseudo-H atom at the Zn-Mg terminal. Generally, the unpassivated Zn$_{1-x}$Mg$_x$O (001) surface is unstable and the surface adsorption or reconstruction is inevitable~\cite{Kresse_PRL2003,Dulub_PRL2003,Valtiner_PEL2009,Goniakowski_RepRrogPhys2008,Lauritsen_ACSNano2011,Meyer_PRB2004,Calzolari_SurfSci2013}. Thus, the surface adsorption and defects are considered to simulate the Zn$_{1-x}$Mg$_x$O/ZnO heterostructures with thin Zn$_{1-x}$Mg$_x$O layers~\cite{Dulub_PRL2003}. As an example, in Fig.~\ref{geome-struc-ZnMgo-ZnO}(a) we give the structure diagram of a Zn$_{1-x}$Mg$_x$O/ZnO heterostructure with two surfaces passivated by pseudo-H atoms. The heterostructure contains a $2\times2$ in-plane (001) Zn$_{0.75}$Mg$_{0.25}$O/ZnO supercell and 18 Zn-Mg-O layers and 18 Zn-O layers. A 15-$\text{\AA}$-thick vacuum layer is added along the [001] direction to prevent any unintentional interactions between the slabs. From the interface to surface, the atomic layers on the ZnO side are labeled as L$\bar{1}$, L$\bar{2}$, $\cdots$, L$\bar{17}$, and L$\bar{18}$, while the atomic layers on the Zn$_{0.75}$Mg$_{0.25}$O side are labeled as L1, L2, $\cdots$, L17, and L18, respectively. The top view of Fig.~\ref{geome-struc-ZnMgo-ZnO}(a) along the [001] direction is shown in Fig.~\ref{geome-struc-ZnMgo-ZnO}(b). Three adsorption sites named On-top, Fcc-hollow, and Hcp-hollow, are indicated by the arrows. The positions of zinc atoms in each layer are numbered as 1, 2, 3, and 4, respectively. For the Mg doping level $x=0.25$ case, the zinc atoms at position 1 are substituted by magnesium atoms in the odd layers, while the zinc atoms at position 3 are replaced in the even layers. For the $x=0.50$ situation, the zinc atoms at positions 2 and 4 are replaced by magnesium atoms in each layer. All calculations are carried out in framework of density
functional theory using the Vienna\emph{ab initio} Simulation Package (VASP)~\cite{calculating_parameter}. The in-plane lattice constants of Zn$_{1-x}$Mg$_{x}$O/ZnO ($x=0.25$ and 0.50) heterostructures are fixed to those of ZnO during the calculations.

\begin{figure}
\includegraphics[scale=1]{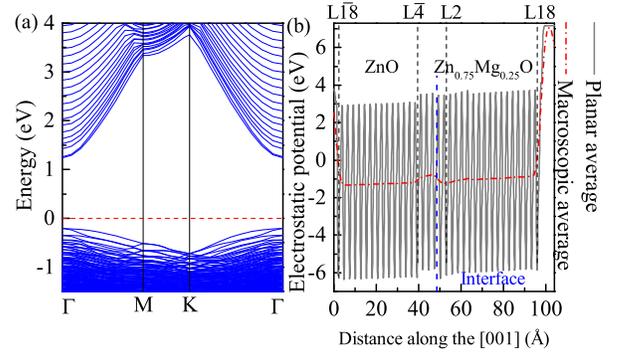}
\caption{(a) The energy band structure of the Zn$_{0.75}$Mg$_{0.25}$O/ZnO heterostructure without oxygen vacancies and with two pseudo-H-atoms-passivated surfaces. (b) The plane average (solid curve) and macroscopic average
(dash-dot curve) electrostatic potential (seen by electron) across the Zn$_{0.75}$Mg$_{0.25}$O/ZnO heterostructure along the [001] direction.}\label{pure-band-pot}
\end{figure}

Figure~\ref{pure-band-pot}(a) shows the band structure of Zn$_{0.75}$Mg$_{0.25}$O/ZnO heterostructure shown in Fig.~\ref{geome-struc-ZnMgo-ZnO}(a) (i.e., the heterostructure has 18 Zn-O and 18 Zn-Mg-O layers and two pseudo-H-atoms-passivated surfaces). Clearly, the valence band maximum (VBM) and the conduction band minimum (CBM) are both located at the $\Gamma$ point, and the Fermi level lies in the band gap. Thus the energy band of the Zn$_{0.75}$Mg$_{0.25}$O/ZnO heterostructure exhibits direct-gap semiconductor characteristics (the calculated  band gap is 1.45\,eV) and no 2DEG is formed at the interface. For the $x=0.50$ case, the band structure is similar to that of the $x=0.25$ and the calculated bad gap is 1.56\,eV. Therefore, 2DEG cannot appear near the interfaces of the perfect Zn$_{1-x}$Mg$_{x}$O/ZnO ($x=0.25$ and 0.50) heterostructures (without defects) with thick Zn$_{1-x}$Mg$_{x}$O layers. We also calculated the electrostatic potential distribution for the above heterostructures, and Fig.~\ref{pure-band-pot}(b) presents the results for the $x=0.25$ case as an example. There is a conspicuous bulge in the macroscopic average potential curve near the interface (from the L$\bar{4}$ Zn-O layer to the L2 Zn-Mg-O layer). In the atomic layers away from the interface, e.g., the Zn-O layers from  L$\bar{4}$ to L$\bar{18}$ or Zn-Mg-O layers from  L2 to L18, the average potential almost retains a constant. Thus, a potential barrier rather than a quantum well is formed near the interface of the Zn$_{0.75}$Mg$_{0.25}$O/ZnO heterostructure. Similar phenomena are also observed in the macroscopic average potential curve of the Zn$_{0.5}$Mg$_{0.5}$O/ZnO heterostructure. This potential barrier should be caused by the polar discontinuity at the interface, which could induce a localized polarization field near the interface. The polarization field cannot cause the bottom of the conduction band to overlap with the top of the valence band as in the case of LaAlO$_3$/SrTiO$_3$ heterostructures~\cite{Lee_PRB2008}. Thus, the polar discontinuity alone cannot explain the observed 2DEG near the interface of Zn$_{1-x}$Mg$_{x}$O/ZnO heterostructure with thick Zn$_{1-x}$Mg$_{x}$O layers.

\begin{figure}
\includegraphics[scale=1]{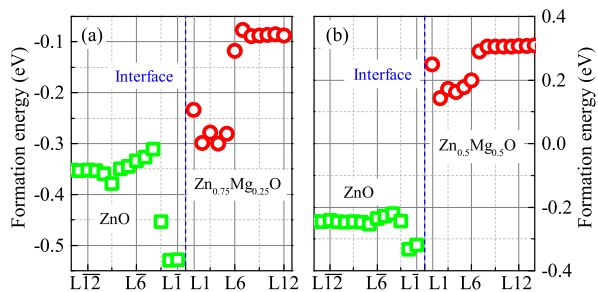}
\caption{Formation energies of oxygen vacancies
at different atomic layers in Zn$_{1-x}$Mg$_{x}$O/ZnO  heterostructures with two pseudo-H-passivated surfaces. (a) For the $x=0.25$, and (b) for the $x=0.50$ heterostructures.}\label{VO-formation-energy}
\end{figure}

Then, why the 2DEGs can be formed in Zn$_{1-x}$Mg$_{x}$O/ZnO heterostructures with thick Zn$_{1-x}$Mg$_{x}$O layers? It should be noticed that as intrinsic defects in ZnO and Zn$_{1-x}$Mg$_{x}$O films, oxygen vacancies are inevitable during device fabrication and could play crucial roles for the formation of 2DEG in Zn$_{1-x}$Mg$_{x}$O/ZnO heterostructures~\cite{Liu_PRB2016,Oba_PRB2008,Clark_PRB2010,Alkauskas_PRB2011}. Next, we investigate the effect of oxygen vacancies on the electronic structures of Zn$_{1-x}$Mg$_{x}$O/ZnO ($x=0.25$ and 0.50) heterostructures with thick Zn$_{1-x}$Mg$_{x}$O layers. First, we calculate the formation energy of oxygen vacancies ($E_{\rm f}$) in each atomic layer of the above heterostructures. In the oxygen-rich limit, $E_{\rm f}$ can be written as~\cite{Li_PRB2011}
\begin{equation}\label{Eq-VO}
E_{\rm f}=E({\rm V_{O}})-(E_{0}-0.5E_{{\rm O_{2}}}),
\end{equation}
where $E({\rm V_{O}})$ and $E_0$ are the calculated total energies of the Zn$_{1-x}$Mg$_{x}$O/ZnO ($x=0.25$ and 0.50) heterostructures with and without oxygen vacancies, and $E_{\rm O_{2}}$ is the calculated total energy of the single O$_{2}$ molecule. For the configuration in Fig.~\ref{geome-struc-ZnMgo-ZnO}, each in-plane supercell contains four oxygen atoms, whose positions are labeled as a, b, c, and d, respectively. The oxygen atoms at d position are removed in a certain fixed layer to create oxygen vacancies in the calculations. Figure~\ref{VO-formation-energy} shows the formation energies of the oxygen vacancies in each layer of the Zn$_{0.75}$Mg$_{0.25}$O/ZnO and Zn$_{0.5}$Mg$_{0.5}$O/ZnO heterostructures with 18 Zn-O and 18 Zn-Mg-O layers, and two pseudo-H-passivated surfaces. Inspection of Fig.~\ref{VO-formation-energy} indicates that the overall variation trends of the $E_{\rm f}$ vs layer number curves for the two heterostructures are similar. Thus we only discuss the variation of $E_{\rm f}$ in the Zn$_{0.75}$Mg$_{0.25}$O/ZnO heterostructure. On the ZnO side, the value $E_{\rm f}$ keeps as a constant in the first two layers, and then sharply increases with increasing layer number, reaches its maximum at L$\bar{4}$, then decreases with further increasing layer number, and  tends to be saturated as the layer number is greater than 9. On the Zn$_{0.75}$Mg$_{0.25}$O side, the values of $E_{\rm f}$ near the interface (L1 to L6 Zn-Mg-O layers) vary between $-0.3$\,eV and $-0.1$\,eV, while those for the layers with layer number being greater than 6 are almost fixed at $-0.1$\,eV. Obviously, the oxygen vacancies can be easily formed on the ZnO side, especially in the first two Zn-O layers near the interface.

\begin{figure}[hb]
\includegraphics[scale=1]{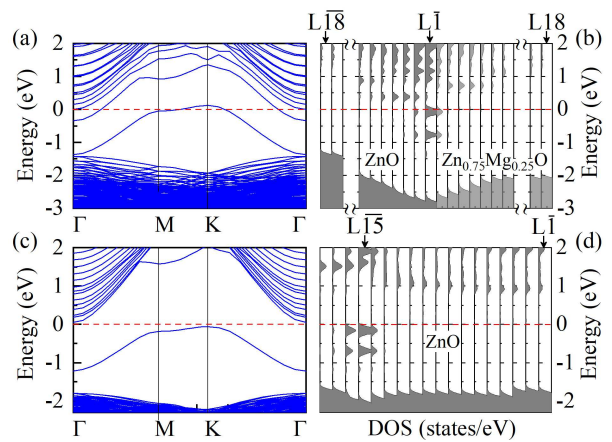}
\caption{(a) The band structure of Zn$_{0.75}$Mg$_{0.25}$O/ZnO heterostructure with oxygen vacancies in the L$\bar{1}$ Zn-O layer. (b) The partial DOS projected onto atomic planes for the  Zn$_{0.75}$Mg$_{0.25}$O/ZnO heterostructure with oxygen vacancies in the L$\bar{1}$ Zn-O layer. (c) The band structure of Zn$_{0.75}$Mg$_{0.25}$O/ZnO heterostructure with oxygen vacancies in the L$\bar{15}$ Zn-O layer. (d) The partial DOS projected onto atomic planes for the  Zn$_{0.75}$Mg$_{0.25}$O/ZnO heterostructure with oxygen vacancies in the L$\bar{15}$ Zn-O layer.}\label{VO-band}
\end{figure}

Considering the variation trends in electronic structures with V$_{\rm O}$ position for the $x=0.25$ and 0.50 heterostructures with two pseudo-H passivated surfaces are also similar, we only present and discuss the results obtained from the $x=0.25$ ones. We first discuss the case that oxygen vacancies are located at the most easily formed position (L$\bar{1}$  layer). Figure~\ref{VO-band}(a) presents the band structure of this configuration. From this figure, one can see that the oxygen vacancies in the L$\bar{1}$ Zn-O layer introduce a defect band in the band gap and the top of the defect band is higher than the Fermi level. At the same time, the Fermi level enters into the bottom of the conduction band, i.e., the conduction band overlaps with the defect band. Thus, part of the electrons in the defect band would be transferred into the conduction band and become conduction electrons. Figure~\ref{VO-band}(b) shows the partial density of states (DOS) projected onto atomic planes for the $x=0.25$ heterostructure with oxygen vacancies in the L$\bar{1}$ Zn-O layer and two pseudo-H passivated surfaces. Clearly, only in L$\bar{2}$ , L$\bar{1}$, and L1 layers the DOS near the Fermi level is nonzero, i.e., the conduction electrons are concentrated in the two Zn-O layers and one Zn-Mg-O layer near the interface. These three layers occupy a space with thickness $\sim$8.4\,{\AA}, which indicates that the 2DEG is formed near the interface of the heterostructure. From the orbital DOS of L$\bar{2}$ to L1 layers, it is found that these conduction electrons are mainly composed of  Zn-4$s$ and O-2$p$ orbitals (not shown). In addition, it is found that when the oxygen vacancies are located in the L$\bar{2}$ and L$\bar{3}$ Zn-O layers and the L1 to L6 Zn-Mg-O layers, their band structures are similar to that in Fig.~\ref{VO-band}(a). However, the band structures of the heterostructures would reveal semiconductor characteristics when the oxygen vacancies are far from the interface (i.e., behind the L$\bar{3}$ Zn-O layer and L6 Zn-Mg-O layer). We take the Zn$_{0.75}$Mg$_{0.25}$O/ZnO heterostructure with oxygen vacancies in the L$\bar{15}$ Zn-O layer as an example. Figure~\ref{VO-band}(c) shows the band structure of this configuration. The oxygen vacancies in the L$\bar{15}$ Zn-O layer also introduce a defect band in the gap, while the top of the defect band is located at 0.11\,eV below the bottom of the conduction band. The Fermi level lies between the conduction band and the defect band. Therefore, the introduction of oxygen vacancies in the L$\bar{15}$ Zn-O layer cannot induce 2DEG at the interface of the heterostructure. Figure~\ref{VO-band}(d) shows the partial DOS projected onto atomic planes for the Zn$_{0.75}$Mg$_{0.25}$O/ZnO heterostructure with oxygen vacancies in the L$\bar{15}$ Zn-O layer. From this figure, one can see that the defect band of the oxygen vacancies is in fact composed of a large number of deep energy levels as far as the energy band of the inner atomic layer is concerned. These deep levels cannot overlap with the conduction band even if the bottom conduction band of the Zn-O layer near the interface is lower than that of the inner Zn-O layer. On the contrary, the defect levels of the oxygen vacancies near the interface layers are so shallow that the bottom of the conduction band overlaps with the top of the defect band [see Fig.~\ref{VO-band}(b)]. This is why 2DEG exists only when the oxygen vacancies are located near the interface of the heterostructure. On the other hand, the defect band formed by oxygen vacancies of inner Zn-O layers could enhance the conductivity of the heterostructure: the device will exhibit a thermal-activated form conductance with activation energy $E_a$, where $E_a$ is about half of the energy difference between the bottom of conduction band and the top of the defect band. Summarizing the results mentioned above, one can readily conclude that the oxygen vacancies near the interface are the origin of the 2DEGs in Zn$_{1-x}$Mg$_{x}$O/ZnO ($x=0.25$ and 0.50) heterostructures with thick Zn$_{1-x}$Mg$_{x}$O layers.

\begin{figure}
\includegraphics[scale=1]{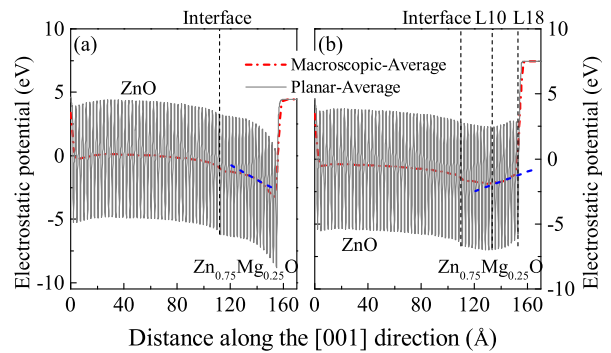}
\caption{(a) The plane average (solid curve) and macroscopic average
(dash-dot curve) electrostatic potential (seen by electron) along [001] direction for the Zn$_{0.75}$Mg$_{0.25}$O/ZnO heterostructure, in which only the surface of ZnO film is passivated by pseudo-H atoms. (b) The plane average (solid curve) and macroscopic average (dash-dot curve) electrostatic potential (seen by electron) along [001] direction of the Zn$_{0.75}$Mg$_{0.25}$O/ZnO heterostructure with a Zn$_{0.75}$Mg$_{0.25}$O surface of H-atom adsorption.}\label{pot}
\end{figure}

Now, we study the origin of 2DEGs in Zn$_{1-x}$Mg$_{x}$O/ZnO ($x=0.25$ and 0.50) heterostructures when the Zn$_{1-x}$Mg$_{x}$O films are very thin. In this situation, we calculate the electronic structures of the heterostructures with 42 Zn-O and 18 Zn-Mg-O layers, in which the surface of the Zn$_{1-x}$Mg$_{x}$O film is no longer passivated. The reason for choosing 42 Zn-O layers (instead of 18 layers) is to obtain the distribution range of 2DEG on the ZnO side. Since the results obtained from the $x=0.25$ and 0.50 heterostructures are also similar, we only present and discuss the results for the $x=0.25$ heterostructure. Figure~\ref{pot}(a) shows the electrostatic potential of Zn$_{0.75}$Mg$_{0.25}$O/ZnO heterostructure (with 42 Zn-O and 18 Zn-Mg-O layers) in which only the surface of ZnO film is passivated by pseudo-H atoms. Obviously, the macroscopic average potential on the ZnO side is insensitive to the position, while it decreases with increasing distance to the interface on the Zn$_{0.75}$Mg$_{0.25}$O side. A macroscopic field perpendicular to the surface with the magnitude of 0.06\,V/{\AA} is obtained by linear fitting the macroscopic average electrostatic potential. This kind of field or potential would lead to an instability of the (001) polar (so-called Tasker type III) surface~\cite{Lauritsen_ACSNano2011,Tasker_JPC1979}. In the light of recent experimental and theoretical results, the polar oxide surfaces can be stabilized via charge transfer between the upper and lower surfaces~\cite{Kresse_PRL2003,Dulub_PRL2003}, adsorption of external atoms~\cite{Kresse_PRL2003,Dulub_PRL2003,Valtiner_PEL2009,Goniakowski_RepRrogPhys2008,Lauritsen_ACSNano2011}, and stoichiometry variations~\cite{Kresse_PRL2003,Dulub_PRL2003,Valtiner_PEL2009,Goniakowski_RepRrogPhys2008}. For Zn$_{1-x}$Mg$_{x}$O/ZnO heterostructures, we consider the effects of adsorption (hydrogen atoms, OH groups, and oxygen atoms) and stoichiometry variations (defects) on the electronic structures of Zn$_{1-x}$Mg$_{x}$O/ZnO ($x=0.25$ and 0.50) heterostructures. Through structural relaxations, it is found that the hydrogen atoms prefer to be adsorbed atop the zinc atom (On-top site), while the preferred adsorption sites for the OH groups and oxygen atoms are the Fcc-hollow sites [see Fig.~\ref{geome-struc-ZnMgo-ZnO}(b)]. Our results are consistent with those in Refs.~\cite{Kresse_PRL2003,Dulub_PRL2003,Valtiner_PEL2009}. For the $2\times 2$ in-plane (001) Zn$_{1-x}$Mg$_{x}$O supercell, the numbers of the  On-top and Fcc-hollow sites are both 4. In our calculations, the coverages of hydrogen atoms, OH groups, and oxygen atoms adsorbed on the surface of Zn$_{1-x}$Mg$_{x}$O are 50\%, 50\%, and, 25\%, respectively, while the concentration of vacancies on the Zn or Mg sites is 25\%~\cite{Kresse_PRL2003,Dulub_PRL2003,Valtiner_PEL2009}. Specifically, for the $x=0.25$ heterostructure, the absorption sites of the hydrogen atoms are set on the top of the zinc atoms at positions 1 and 4 [see Fig.~\ref{geome-struc-ZnMgo-ZnO}(b)]; the adsorption sites for the OH groups are set at Fcc-hollow positions located at the top of the arrow and the position of the black dot; the Fcc-hollow position at the top of the arrow is also set as the adsorption site of oxygen atoms; the Zn vacancies are obtained via removing the zinc atoms located at position 1.

Figure~\ref{pot}(b) shows the electrostatic potential of Zn$_{0.75}$Mg$_{0.25}$O/ZnO heterostructures with hydrogen atoms adsorbed on the Zn$_{0.75}$Mg$_{0.25}$O surface (and with 42 Zn-O and 18 Zn-Mg-O layers). The results for the Zn$_{0.75}$Mg$_{0.25}$O surface with oxygen atoms adsorption, OH groups adsorption, and Zn or Mg vacancies are similar to that shown in Fig.~\ref{pot}(b). The macroscopic average potential on the ZnO side remains nearly a constant after adsorption of hydrogen atoms. On the Zn$_{0.75}$Mg$_{0.25}$O side, the macroscopic average potential is almost insensitive to the position from L1 to L9 layers, and then slightly increases with increasing distance to the interface.  An electrostatic field (with magnitude of $\sim$0.038\,V/{\AA}) being opposite to that shown in Fig~\ref{pot}(b) exists between L10 and L18 layers. Thus surface adsorption or metal ion vacancies could really stabilize the polar surfaces of the Zn$_{1-x}$Mg$_{x}$O/ZnO ($x=0.25$ and 0.50) heterostructures.

\begin{figure}
\includegraphics[scale=1]{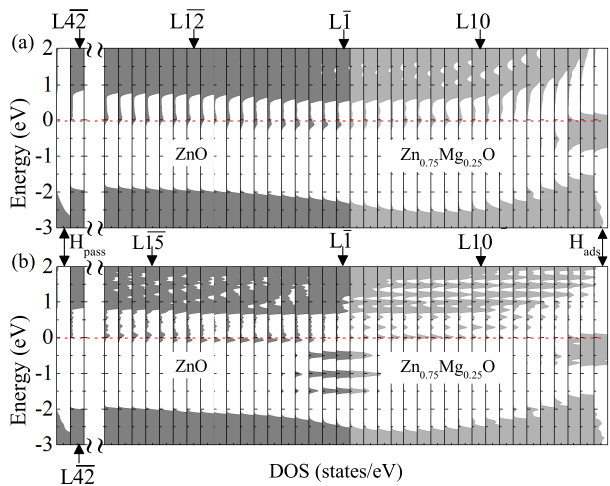}
\caption{The partial DOS projected onto the atomic layers for the Zn$_{0.75}$Mg$_{0.25}$O/ZnO heterostructures with surfaces of (a) H-atom absorption, and (b) O-atom absorption.}\label{LDOS}
\end{figure}

The electronic structures of the Zn$_{1-x}$Mg$_{x}$O/ZnO ($x=0.25$ and 0.50) heterostructures with exotic-atoms-adsorbed surfaces or with surfaces having metal ion vacancies, have been also calculated. It is found that the electronic structures of the heterostructures reveal semiconductor characteristics when the surface contains metal ion vacancies, while the electronic structure exhibits metallic characteristics as the hydrogen, oxygen atoms, and OH groups are adsorbed on the surface, respectively. Figure~\ref{LDOS}(a) show the partial DOS decomposed to the atomic layers for the Zn$_{0.75}$Mg$_{0.25}$O/ZnO heterostructures (42 Zn-O and 18 Zn-Mg-O layers) with hydrogen atoms adsorption. For the oxygen-atom- or OH-groups-adsorption case, the partial DOS plot is similar to that in Fig.~\ref{LDOS}(a).  Clearly, the adsorption of hydrogen atoms on the surface introduces defect states in the gap. Although polarization field distributed in the L10 to L18 Zn-Mg-O layers has significantly lifted up the top of the valence band, the valence band is still far from overlapping with the conduction band.  However, the defect states introduced by hydrogen atoms are located near the Fermi level, and partially higher than the Fermi level. As a result, the defect band overlaps with the conduction band, which renders the heterostructure to exhibit metallic characteristics in electronic structures. Inspection of Fig.~\ref{LDOS}(a) also indicates that the conduction electrons are distributed from L$\bar{12}$ to L10 layers, i.e., in the range of $\sim$5.53\,nm near the interface. Thus, the heterostructure would reveal 2D or quasi-2D behaviors in transport properties. We also calculated the electronic structures of the Zn$_{1-x}$Mg$_{x}$O/ZnO ($x=0.25$ and 0.50) heterostructures with oxygen vacancies and surfaces adsorbed with exotic atoms. It is found that the introduction of an oxygen vacancy in a certain atomic layer of the 2$\times$2 in-plane supercell would also produce a defect band in the gap. However, the defect band does not overlap with the conduction band, i.e., the introduction of oxygen vacancies does not change the ultimate properties of the heterostructures. As an example, in Fig.~\ref{LDOS}(b) we give the partial DOS projected onto the atomic layers for the H-adsorbed Zn$_{0.75}$Mg$_{0.25}$O/ZnO heterostructure (with 42 Zn-O and 18 Zn-Mg-O layers) with oxygen vacancies in L$\bar{1}$. Comparing the partial DOS of the heterostructure without oxygen vacancies [Fig.~\ref{LDOS}(a)], one can see that oxygen vacancies in L$\bar{1}$ introduce an extra defect band, whose maximum is located at 0.15\,eV below the bottom of the conduction band. In this situation, the 2DEG distributes from L$\bar{15}$ to L10 layers ($\sim$6.53\,nm) and still originates from the adsorption of hydrogen atoms. The electronic structure of the heterostructures with oxygen-atoms-adsorbed or OH-groups-adsorbed surface is similar to that for the H-adsorbed heterostructure. In addition, the introduction of oxygen vacancies in Zn-O layer near the interface does not change the semiconductor characteristic of the electronic structure for the heterostructure with metal ion vacancies in the surface of Zn$_{1-x}$Mg$_{x}$O film. Thus, for Zn$_{0.75}$Mg$_{0.25}$O/ZnO heterostructure with thin Zn$_{1-x}$Mg$_{x}$O film, the adsorption of hydrogen atoms, oxygen atoms, or OH groups on the surface of Zn$_{1-x}$Mg$_{x}$O layer is responsible for the formation of 2DEG near the interface.

In summary, to explore the origin of 2DEGs in Zn$_{1-x}$Mg$_{x}$O/ZnO heterostructures, we constructed the Zn$_{1-x}$Mg$_{x}$O/ZnO ($x=0.25$ and 0.50) heterostructures with different surfaces and investigated their electronic structures by first-principles calculations. It is found that the polarity discontinuity near the interface can neither lead to the formation of 2DEGs in devices with thick Zn$_{1-x}$Mg$_{x}$O layers nor in devices with thin Zn$_{1-x}$Mg$_{x}$O layers. For the heterostructures with thick Zn$_{1-x}$Mg$_{x}$O layers, the oxygen vacancies near the interface are the source of the 2DEGs. For the heterostructures with thin Zn$_{1-x}$Mg$_{x}$O layers, adsorption of hydrogen atoms, oxygen atoms, or OH groups on the surface of Zn$_{1-x}$Mg$_{x}$O films can not only stabilize the polar surface of Zn$_{1-x}$Mg$_{x}$O layer, but also cause the formation of 2DEGs near the interfaces of the devices.

%\begin{acknowledgments}
The calculation was conducted on the CJQS-HPC platform at Tianjin University. This work is supported by the National
Natural Science Foundation of China through Grants No. 12174282.
%\end{acknowledgments}

\end{document}